\documentstyle[12pt]{article}
\title{ {\normalsize{{\hskip 9cm} BIHEP-TH-95-27, Oct 1995}}\\
        CP-Violation For $B \rightarrow X_sl^+l^-$ Including
        Long-Distance Effects}
\author{ Dong-Sheng Du$\rm ^{a,b}$, Mao-Zhi Yang$\rm ^b$ 
         \thanks{E-mail adress: Duds@bepc3.ihep.ac.cn;$~~$
         Yangmz@bepc3.ihep.ac.cn} \\
         $\rm ^a$ \it{CCAST(World Laboratory), 
         ~~P.O.Box 8730, Beijing 100080, China}\\
         $\rm ^b$ \it{Institute of High Energy Physics, Chinese 
         Academy of Sciences,}\\
         \it{P.O.Box 918(4), Beijing, 100039, P. R. China} 
         \thanks{mailing address.}}

\date{}
\begin{document}
\maketitle
\vspace*{0.3cm}
\begin{flushleft}
\end{flushleft}

\unitlength=1cm
\begin{flushleft}
\begin{picture}(16, 0.2)(0, 0)
\put(0, 0){\line(1, 0){16}}
\end{picture}
\section*{Abstract}
\end{flushleft}

We consider the $CP$ violating effect for $B\to X_sl^+l^-$ process, including
both short and long distance effects. We obtain the $CP$ asymmetry parameter
and present its variation over the dilepton mass.
\begin{flushleft}
\begin{picture}(16, 0.2)(0, 0)
\put(0, 0){\line(1, 0){16}}
\end{picture}
\end{flushleft}

\newpage

As well known, the flavor changing process $b\to sl^+l^-$ can serve as an 
excellent ``window" for precisely testing the standard model or for finding 
new
physics beyond it. This process occurs through the one-loop diagrams. There
are three types of Feynman diagrams for $b\to sl^+l^-$ transition, they are
electromagnetic (photonic) penguin diagrams, weak ($Z^0$ boson) penguin
diagrams, and box diagrams [1,2]. These diagrams produce the short distance
contributions to this process. The short distance contribution to 
the branching ratio of the inclusive process $B\to X_sl^+l^-$ is estimated
to be about $10^{-5}$ at large mass of top quark [2,3]. In addition to the 
short distance contributions, there are long distance contributions to
$b\to sl^+l^-$ through physical intermediate states:
    $$b\to s(u{\bar u},c {\bar c})\to sl^+l^-.$$
The intermediate states can be vector mesons such as 
$\rho$, $\omega$, $J/\psi$
, and $\psi'$. The long distance contribution to the branching ratio of 
$b\to sl^+l^-$ is calculated to be as large as $10^{-3}$ [4,5]. So the
long distance effect is not negligable. In this paper, we study the long 
distance effect in the $CP$-violation of the inclusive process 
$B\to X_sl^+l^-$. Our work is different from previous ones in two aspects. 
Firstly, in
Ref.[1], the authors studied the CP-violation effect of $B\to X_sl^+l^-$
by considering only photonic penguin diagrams, here, we consider all the
three types of the diagrams (electromagnetic, weak, and box diagrams) 
and include QCD corrections within the leading 
logarithmic approximation [6]. Secondly, we consider both short and long 
distance contributions.   

The effective Hamiltonian relevant to $b\to sl^+l^-$ transitions is [3,6,7,8]
$$\begin{array}{rl}
\displaystyle{\cal H}_{eff}=\frac{G_F}{\sqrt{2}}\left(
      \frac{\alpha}{4\pi S_w^2}\right)
         \displaystyle\sum_i V_i&[A_i{\bar s}\gamma_\mu(1-\gamma_5)b{\bar l}
         \gamma^{\mu}(1-\gamma_5)l+\\
         &+B_i{\bar s}\gamma_{\mu}(1-\gamma_5)b{\bar l}
         \gamma^\mu(1+\gamma_5)l\\
         &-2im_bS_w^2F_2^i{\bar s}\sigma_{\mu\nu}q^{\nu}(1+\gamma_5)/q^2
          b{\bar l}\gamma^\mu\gamma_5l],
\end{array} \eqno(1)$$
where $V_i=U_{is}^*U_{ib}$ ($i=u,c,t$) is the product of CKM matrix elements.
$S_w=sin\theta_w$,
$\theta_w$ is the Weinberg angle. $l=e$, $\mu$. q is the momentum of 
the lepton pair.

At the scale $\mu\approx M_w$, the coefficients $A_t$ and $B_t$ take the forms:
$$\begin{array}{rl}
A_t&=~-2B(x)+2C(x)-S_w^2[4C(x)+D(x)-4/9]\\
B_t&=~-S_w^2[4C(x)+D(x)-4/9]
\end{array} \eqno(2)
$$
where $x=m^2_t/M_w^2$.
$$
\begin{array}{rl}
B(x)&=~\frac{1}{4}\left[\displaystyle\frac{-x}{x-1}+
      \displaystyle\frac{x}{(x-1)^2}lnx\right]\\[4mm]
C(x)&=~\displaystyle\frac{x}{4}\left[\displaystyle\frac{x/2-3}{x-1}+
        \displaystyle\frac{3x/2+1}{(x-1)^2}lnx\right]\\[4mm]
D(x)&=~\left[\displaystyle\frac{-19x^3/36+25x^2/36}{(x-1)^3}+
      \displaystyle\frac{-x^4/6+5x^3/3-3x^2+16x/9-4/9}{(x-1)^4}lnx\right].
\end{array} \eqno(3)
$$
Here, $B(x)$ arises from box diagram, and $C(x)$ from $Z^0$ penguin diagram,
while $D(x)$ is contributed from $\gamma$ penguin diagram. We can see from
eq.(3) that with $x$ coming larger, the contribution from box diagram and
$\gamma$ penguin diagram will decline, while $C(x)$ will become dominant.
Then using the renormalization group equation to scale the effective
Hamiltonian down to the order of 
the $b$ quark mass, one obtain
$$\begin{array}{rl}
A_t(x,\xi)&=~A_t(x)+\displaystyle\frac{4\pi}{\alpha_s(M_w)}\left\{
            -\displaystyle\frac{4}{33}(1-\xi^{-11/23})+
            \displaystyle\frac{8}{87}(1-\xi^{-29/23})\right\}S_w^2\\
B_t(x,\xi)&=~B_t(x)+\displaystyle\frac{4\pi}{\alpha_s(M_w)}\left\{
            -\displaystyle\frac{4}{33}(1-\xi^{-11/23})+
            \displaystyle\frac{8}{87}(1-\xi^{-29/23})\right\}S_w^2 
\end{array} \eqno(4)
$$
where $\xi=\frac{\alpha_s(m_b)}{\alpha_s(M_w)}=1.75.$

Moreover, the coefficient for the magnetic-moment operator is given by
$$\begin{array}{rl}
F_2^t(x,\xi)~=&\xi^{-16/23}\left[-\displaystyle\frac{1}{12}
            \displaystyle\frac{8x^3+5x^2-7x}{(x-1)^3}
           +\displaystyle\frac{3x^3/2-x^2}{(x-1)^4}lnx\right.\\[4mm]
          & \left.-\displaystyle\frac{116}{135}(\xi^{10/23}-1)
           -\displaystyle\frac{58}{189}(\xi^{28/23}-1)\right].
\end{array}
\eqno(5)$$

In our numerical calculation, we take $m_t=174GeV$ [9]. Furthermore the 
non-resonant coefficients $A_i$, $B_i$ (i=u,c) are represented by
$$A_i=B_i=a_2 S_w^2g\left(\frac{m_i^2}{m_b^2},\frac{q^2}{m_b^2}\right),
\eqno(16)                        
$$
with [7]
$$g(r_i,s)=\left\{ 
\begin{array}{ll}
\frac{4}{3}lnr_i-\frac{8}{9}-\frac{4}{3}\frac{4r_i}{s}+\frac{2}{3}\sqrt{1
  -\frac{4r_i}{s}}(2+\frac{4r_i}{s})(ln\displaystyle\frac{1+\sqrt{1-4r_i/s}}
  {1-\sqrt{1-4r_i/s}}+i\pi),  &\mbox (\frac{4r_i}{s}<1);\\
\frac{4}{3}lnr_i-\frac{8}{9}-\frac{4}{3}\frac{4r_i}{s}+\frac{4}{3}\sqrt{
  \frac{4r_i}{s}-1}(2+\frac{4r_i}{s})arctan\displaystyle\frac{1}
  {\sqrt{4r_i/s-1}},  &\mbox (\frac{4r_i}{s}>1).
\end{array}
\right.
\eqno(7)
$$
Here $a_2=C_-+C_+/3$, is the coupling for the neutral $b{\bar s}q {\bar q}$
($q=u,c$) four-quark operator.

In addition to the short distance contribution, the inclusive decay 
$B\to X_sl^+l^-$ involves the long distance contributions arising from
$u{\bar u}$ and $c {\bar c}$ resonances, such as $\rho(770)$, $\omega(782)$,
$J/\psi(3100)$, and $\psi'(3700)$ et.c. The long distance contribution to the
coefficients A and B in Eq.(1) can be taken as [4,5,10,11]
$$A_v=B_v=\frac{16\pi^2}{3}\left(\frac{f_v}{M_v}\right)^2
          \frac{a_2 S_w^2}{q^2-M_v^2+iM_v\Gamma_v}e^{2i\phi_v} .\eqno(8)$$
where $M_v$, and $\Gamma_v$ are the mass and width of the relevant
vector meson $\rho$, $\omega$, $J/\psi$, and $\psi'$, respectively.
$e^{2i\phi}$ is the relevant phase between the resonant and non-resonant
amplitude. The decay constant $f_v$ is defined as
$$<0|{\bar c}\gamma_{\mu} c|V(\epsilon)>=f_v\epsilon_{\mu}. \eqno(9)$$
We can determine $f_v$ through the measured partial width for the decays of
the mesons to lepton pairs [12],
$$\Gamma(v\to l^+l^-)=\frac{4\pi}{3}\frac{(Q_c \alpha)^2}{M_v^3}f_v^2,
  \eqno(10)$$
with $Q_c=\frac{2}{3}$. For the parameter $a_2$, there is the CLEO data
$|a_2|=0.26\pm 0.03$ [15]. In this work, $a_2$ should be taken as 
$a_2=-(0.26\pm 0.03)$, and $\phi_v=0$ or $a_2=0.26\pm 0.03$, 
$\phi_v=\frac{\pi}{2}$ [11].

The differential decay width of the inclusive process $B\to X_sl^+l^-$ over
the dilepton mass is given by [4]
$$\frac{d}{dz} \Gamma(B\to X_sl^+l^-)=\frac{G_F^2m_b^5}{192\pi^3}
       \left[\frac{\alpha}{4\pi S_w^2}\right]^2F_b(z), \eqno(11)$$
where $z=\frac{q^2}{m_b^2}$.
$$\begin{array}{rl}
F_b(z)~=&[|V_i A_i(z)|^2+|V_iB_i(z)|^2]f_1^b(z)+\\
       &+S_w^2\{V_i^*V_j[A_i(z)+B_i(z)]^*F_2^j+H.C.\}f_{12}^b(z)+\\
       &+2 S_w^4|V_iF_2^i|^2f_2^b(z), 
\end{array} \eqno(12)$$
and
$$\begin{array}{rl}
f_1^b(z)&=~2(1-z)(1+z-2z^2)\\
f_{12}^b(z)&=~6(1-z)^2\\
f_2^b(z)&=~4(1-z)(1/z-\frac{1}{2}-z/2).
\end{array} \eqno(13)$$

We define the $CP$-violating asymmetry through the rate difference between
$B$ and ${\bar B}$:
$${\cal A}_{cp}=\frac{\Gamma_{{\bar b}}-\Gamma_b}{\Gamma_{{\bar b}}+\Gamma_b}
\eqno(14) $$
where $\Gamma_b$ is obtained by integrating Eq.(11) over the dilepton mass
squared $z$ from $z_{min}=(\frac{2m_l}{m_b})^2$ to 
$z_{max}=(1-\frac{m_s}{m_b})^2$.   The CKM matrix in Eq.(12) can be written 
in terms of four parametres $\lambda$, $A$, $\rho$ and
$\eta$ in the Wolfenstein parametrization[14]. There have been definite 
results for $\lambda$ and $A$, which are
$\lambda=0.2205\pm 0.0018$ [15] and $A=0.80\pm 0.12$ [16]. But for $\rho$ and
$\eta$, there are not definite results. So we express the CP-violating
parameter for $B\to X_s e^+e^-$ in terms of $\rho$ and $\eta$,
$${\cal A}_{cp}^{S+L}=\frac{7.618\times 10^{-3}\eta}{1.799+
2.991\times 10^{-3}\rho+
45.876(1+0.0484^2\eta^2)+1.4718\times 10^{-4}(\rho^2+\eta^2)} \eqno(15)$$
\noindent for the case of including long distance effects, and
$${\cal A}_{cp}^{S}=\frac{3.1389\times 10^{-3}\eta}{1.702+
3.345\times 10^{-3}\rho+
3.607\times 10^{-2}(1+0.0484^2\eta^2)+
1.4373\times 10^{-4}(\rho^2+\eta^2)} \eqno(16)$$
\noindent for the case without long distance effects. Eq.(15) and Eq.(16)
indicate that $\eta$ affects the CP asymmetry mainly, and $\rho$ does not.

In table I, we give the results of ${\cal A}_{cp}$ for some ``best values"
of $(\rho,~\eta)$ [16]. We can see that,
i) without the long distance effects, the CP-violating asymmetry 
${\cal A}_{cp}$ is about $(1.8\sim6.1)\times 10^{-4}$, 
while, in Ref.[1], the relevant $CP$ asymmetry is
anout $1.3\times 10^{-2}$. Our result is about twenty times smaller than theirs. The reason 
is that, in Ref.[1], only the photonic penguin is considered. But in fact
the $Z^0$ penguin will give big contribution to the amplitude at large
$m_t(\sim 174GeV)$ [2], at the same time, it doesn't provide large 
CP-nonconserving phase,
ii) including the long distance effects, the result of the CP asymmetry 
parameter ${\cal A}_{cp}$ is about $(1.5\sim 5.4)\times 10^{-5}$. It is 
reduced about one order by the resonant effects. The main difference between
the cases with and without long distance effect resides in the third term
of the denominator of eq.(15) and (16), which comes from the integration
of the first term of eq.(12), i.e., 
$\int^{z_{max}}_{z_{min}}dz[|V_c|^2|(A_c(z)+B_c(z))|^2f_1^b(z)]$.
Without resonant contributions
$$\int^{z_{max}}_{z_{min}}dz[|V_c|^2|(A_c(z)+B_c(z))^S|^2f_1^b(z)]=0.12|V_c|^2,
\eqno(17)$$
While with resonant contributions
$$\int^{z_{max}}_{z_{min}}dz[|V_c|^2|(A_c(z)+B_c(z))^{S+L}|^2
f_1^b(z)]=152.9|V_c|^2,
\eqno(18)$$
Because the total decay width of $J/\psi$ or $\psi'$ is narrow 
($\Gamma_{J/\psi}=88KeV$, $\Gamma_{\psi'}=277KeV$ ), when the dilepton
mass squared $z$ is near the mass squared of $J/\psi$ or $\psi'$, the 
resonance will give a big contribution. At the same time, the first term
of eq.(12) only contribute to the decay width $\Gamma_b$ and $\Gamma_{\bar b}$,
it does not give contribution to the CP-violation. So with the resonant
effects, the CP-violation will be reduced greatly.

We also calculated the distribution of the CP asymmetry over the dilepton 
mass for $(\rho,~\eta)$ taking the ``preferred value" of $(-0.12,~0.34)$ [16], 
$$
a_{cp}=\frac{F_{{\bar b}}(z)-F_b(z)}{F_{{\bar b}}(z)+F_b(z)}.
\eqno(19) $$
The result is ploted in Fig.1. The solide line is for the case without
resonances and the dotted line for the case with resonances. We can see 
that, in general, the CP asymmetry is suppressed by the resonance effect,
and in the region near the resonances, CP-violating parameter is suppressed
severely.

E. Golowich and S. Pakvasa have discussed the long range effects in $B\to
K^*\gamma$ [17], which is relevant to the condition of the squared mass
$q^2=0$. They found a small effect by respecting gauge invariance. 
It should be noted that there is no controversy between their results and ours. 
That is in Fig.1, it is shown that when $q^2\to 0$ the long distance effect
is also very small in the case of $B\to X_s l^+l^-$.

Finally, we want to point out that for the case of $l=\mu $, the CP 
asymmetry parameter is smaller than the $l=e$ case.

\vspace*{1cm}
This work is supported in part by the China National Natural 
Science Foundation and the Grant of State Commision of Science and 
Technology of China.

\newpage

\begin{center}
\begin{tabular}{|c|c|c|}
\hline
$(\rho,~\eta)$ & ${\cal A}_{cp}^{S}$ & ${\cal A}_{cp}^{S+L}$\\
\hline
(-0.48, 0.10) & $1.81\times 10^{-4}$ & $ 1.60\times 10^{-5}$\\
\hline
(-0.44, 0.12) & $2.17\times 10^{-4}$ & $ 1.92\times 10^{-5}$\\
\hline
(-0.40, 0.15) & $2.71\times 10^{-4}$ & $ 2.40\times 10^{-5}$\\
\hline
(-0.36, 0.18) & $3.25\times 10^{-4}$ & $ 2.88\times 10^{-5}$\\
\hline
(-0.32, 0.21) & $3.79\times 10^{-4}$ & $ 3.35\times 10^{-5}$\\
\hline
(-0.28, 0.24) & $4.34\times 10^{-4}$ & $ 3.83\times 10^{-5}$\\
\hline
(-0.23, 0.27) & $4.88\times 10^{-4}$ & $ 4.31\times 10^{-5}$\\
\hline
(-0.17, 0.29) & $5.24\times 10^{-4}$ & $ 4.63\times 10^{-5}$\\
\hline
(-0.11, 0.32) & $5.78\times 10^{-4}$ & $ 5.11\times 10^{-5}$\\
\hline
(-0.04, 0.33) & $5.96\times 10^{-4}$ & $ 5.27\times 10^{-5}$\\
\hline
(+0.03, 0.33) & $5.96\times 10^{-4}$ & $ 5.27\times 10^{-5}$\\
\hline
(-0.12, 0.34) & $6.14\times 10^{-4}$ & $ 5.43\times 10^{-5}$\\
\hline
\end{tabular}
\end{center}

\vspace{1cm}

Table I. $CP$ asymmetries for some ``best values" of $(\rho,~\eta)$.
${\cal A}_{cp}^{S}$ denotes the cases without long distance contributions,
${\cal A}_{cp}^{S+L}$ with long distance contributions.

\newpage

\begin{flushleft}
\section*{Figure Captions}
\end{flushleft}

\vspace{1cm}

{\bf Fig. 1.} The dilepton mass distribution of the CP asymmetry parameter for
$B\to X_se^+e^-$ process without resonances (solide line) and with resonances
(dotted line). 

\end{document}